\documentclass[journal=jcim,manuscript=article]{achemso}

\usepackage[version=3]{mhchem} % Formula subscripts using \ce{}
\usepackage{hyperref}       % hyperlinks
\usepackage{url}            % simple URL typesetting
\usepackage{booktabs}       % professional-quality tables
\usepackage{amsfonts}       % blackboard math symbols
\usepackage{nicefrac}       % compact symbols for 1/2, etc.
\usepackage{microtype}      % microtypography
\usepackage{graphicx}
\usepackage{natbib}
\usepackage{doi}
\usepackage{xcolor}
\usepackage{tikz}
\usepackage{amsmath}
\usepackage{amssymb}

\SectionNumbersOn

\author{Ronen Taub}
\affiliation[Technion]{Department of Physiology, Biophysics \& Systems Biology, Medicine Faculty, Technion IIT, Haifa, Israel}
\author{Yonatan Savir}
\affiliation[Technion]{Department of Physiology, Biophysics \& Systems Biology, Medicine Faculty, Technion IIT, Haifa, Israel}
\email{yoni.savir@technion.ac.il}

\title[An \textsf{achemso} demo]
  {SAF: Smart Aggregation Framework for Revealing Atoms Importance Rank and Improving Prediction Rates in Drug Discovery}

\abbreviations{SAF}

\begin{document}

\begin{abstract}
   Machine learning, and representation learning in particular, has the potential to facilitate drug discovery by screening a large chemical space \emph{in silico}. A successful approach for representing molecules is to treat them as a graph and utilize graph neural networks. One of the key limitations of such methods is the necessity to represent compounds with different numbers of atoms, which requires aggregating the atom's information. Common aggregation operators, such
as averaging, result in loss of information at the atom level. In this work, we propose a novel aggregating approach where each atom is weighted non-linearly using the Boltzmann distribution with a hyperparameter analogous to temperature. We show that using this weighted aggregation improves the ability of the gold standard message-passing neural
network to predict antibiotic activity. Moreover, by changing the temperature hyperparameter, our approach can reveal the atoms that are important for activity prediction in a smooth and consistent way, thus providing a novel, regulated attention mechanism for graph neural networks.
We further validate our method by showing that it recapitulates the functional group in $\beta$-Lactam antibiotics. The ability of our approach to rank the atoms' importance for a desired function can be used
within any graph neural network to provide interpretability of the results and predictions at the node level.
\end{abstract}

\section{Introduction}

One of the promising current trajectories in drug discovery is harnessing the power of machine learning to screen for billions of compounds \textit{in silico} \cite{stephenson_survey_2019,ekins_exploiting_2019,zhang_machine_2017,lavecchia_machine-learning_2015,vamathevan_applications_2019}. The main advantage of this approach is that it requires a reasonable experimental effort to produce a training set based only on the order of thousands of compounds. One of the key challenges in applying machine learning for drug discovery is the representation of the molecule. The classic approach is to define a feature vector composed of properties based on chemical properties, such as competition of functional groups \cite{moriwaki_mordred_2018,rogers_extended-connectivity_2010}. Once the molecule is represented as a feature vector, the machine can learn which features are associated with some function. In this approach, the definition of the features representing the molecule requires extensive knowledge and prior knowledge in biochemistry. Representation learning, a newer approach in which the machine learns not only what are the features that are relevant to a function but also the relevant features that represent the data has emerged as a powerful tool \cite{bengio_representation_2014}. 

A successful approach for representing molecules is to consider them as a graph and utilize graph neural networks (GNN) such as graph convolutional network (GCN) \cite{duvenaud_convolutional_2015,lin_kgnn_2020,li_grapher_2020} and message passing neural network (MPNN) \cite{yang_analyzing_2019-1,stokes_deep_2020-1}. GNNs are designed to learn how to effectively cast the atomic structure graphs of observed molecules onto new feature space, generating representative fingerprints that can be fed into common predictors. Compared to other neural networks, like large language models (LLMs), or convolutional neural networks (CNNs), GNNs are better suited for drug discovery for two main reasons: first, it takes into account the atomic structure by design, and, second, it generalizes well, even with limited data for training. In recent years, the implementations of GNNs for drug discovery have been innovated with modifications like graph attention layer (GAT) \cite{velickovic_graph_2018}, pooling layers \cite{islam_mpool_2023}, and graph neural networks self-supervised regulation \cite{ju_tgnn_2023}.

\begin{figure}
    \centering
    \includegraphics[width=16cm,clip]{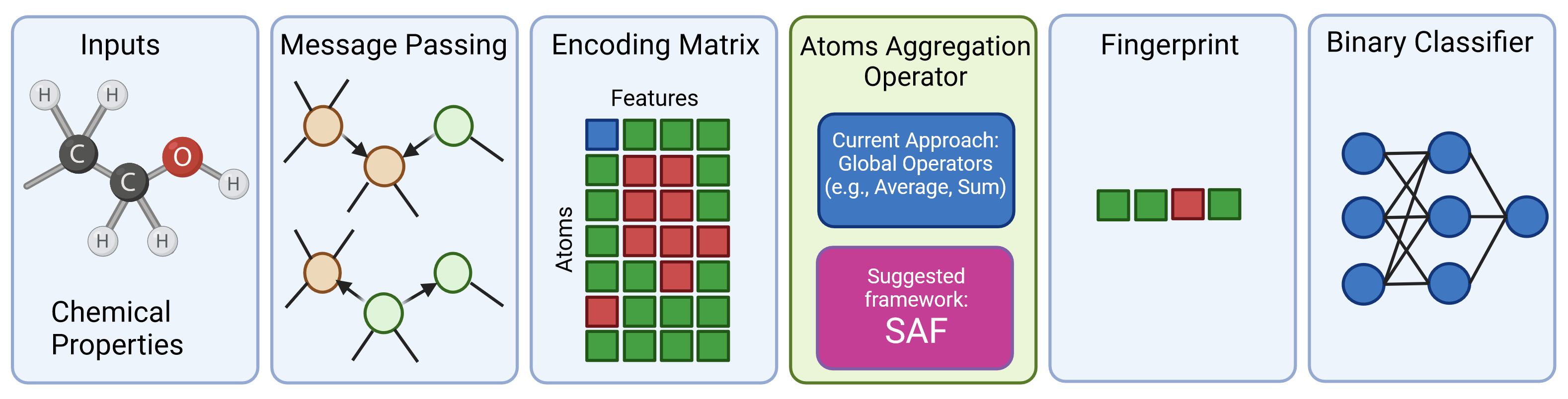}
    \caption{Layout of the main steps of MPNN pipeline. In the message passing phase, the graph is set to be the compound structure such that the nodes and edges are initialized to be the atoms and the bonds. After a number of iterations of the message passing routine, the hidden state of each node is collected and inserted into an encodings matrix (as rows). As different molecules have different number of atoms, an aggregation step is needed. The most common aggregation operator is to average over all the atoms such that the contribution from each atom is the same thus leading to loss of information. Here, the proposed framework, SAF, provides an aggregation scheme that learns the relative contribution of each atom and can be incorporated into any MPNN pipeline. SAF atom ranking provides better classification and interpretability.}
    \label{fig:cartoon}
\end{figure}

There are three main steps in the MPNN pipeline: the encoding step, the aggregation step, and the classification step (Fig. \ref{fig:cartoon}). In this paper, we focus on the aggregation step. The aggregation step consists of an aggregative operator which maps the atoms' encodings $\{\vec{a}_i\}_{i=1}^{N_a}$, where $\vec{a}_i\in \mathrm{R}^{d_h}$ and $N_a$ is the number of atoms, onto 1D fingerprint vector $\vec{f}\in \mathrm{R}^{d_h}$. The aggregation is mandatory as many classifiers, particularly deep learners, are designed to receive fixed-size vectors as input, while compounds have a varying number of atoms $N_a$. The identity of the aggregation operator is a major factor in the model's performance, as it dictates the observability of the encoder's inner states. For example, the most commonly used operator of averaging over all the atoms will only transmit global patterns, traits that all the atoms share. Consequently, the encoder will be biased to create only global patterns, and the binary classifier will be sensitive to global patterns only. 

Moreover, the aggregation step greatly affects the model's interpretability. It is the step where we first lose the distinction capability in the level of the atoms. This significantly limits our ability to interpret which atoms contribute to the prediction of certain properties and to what extent. Methods to interpret GNN-based pipelines at the node level have been researched for quite some time now using two main approaches: reverse engineering of the converged model, for example \cite{ying2019gnnexplainer}, or the use of GAT \cite{velickovic_graph_2018}, which also been modified to support other applications \cite{vinas_hypergraph_2023}. The first approach is less appealing, as there is a profound question mark regarding the relation between the convergence point and the true importance factor of each node. The converged model can be partially degenerative due to limitations in the observed data. Also, there could be conceptually wrong micro-transactions in the message-passing routine with positive effects. The second approach, the use of GAT, is more powerful from our perspective, as it forces the model to decide, which edge it should attend to. In practice, the attention mechanism in GAT, implicitly makes that described interpretability question part of the general objective. Yet, the attention mechanism in GAT adds a substantial amount of new trainable parameters, and therefore it adds ambiguity to the training routine. In certain cases, there can also be conflicts, as there could be nodes that are not important for manifesting the property, but are important for efficient message-passing communications, and vice versa. But more importantly, GAT focuses on ranking edges in the local environment of each node, while neglecting the main question of the importance ranking of each node globally (in the molecule).

In this work, we propose a novel approach coined smart aggregation framework (SAF) method that is free of these problems, thanks to shifting the attention mechanism into a later stage (not interfering with the GNN encoder). Also, the SAF operation incorporates a hyperparameter that allows the user to observe how the attention (importance) scheme transitions between global and local perspectives, in a smooth manner. This offers a more robust and informative view for interpreting and determining the importance of each atom to the learned function. We show that the SAF modification improves antibacterial prediction performance. To further validate our results, we show that in the case of $\beta$-Lactam antibiotics the important atoms converge into the known functional scaffold.

\section{Methods}

\subsection{SAF: Smart Aggregation Framework}\label{subsec:saf}

The starting point of the Smart Aggregation Framework (SAF in short) is the atoms' encodings available from the GNN encoder (or any other molecule encoder). Let $\alpha\in\mathrm{R}^{N_a\times N_f}$ be the encodings matrix containing the atoms encodings $\{\vec{a}_i\}_{i=1}^{N_a}$, so that $\alpha_{ij}$ is the encoding value of the $i^{th}$ atom on the $j^{th}$ feature coordinate (where $N_a$ is the number of atoms, and $N_f$ is the number of features, set to be the hidden state dimension $d_h$ of the encoder). 

The objective of this framework is to sum the elements in $\alpha$ over all the atoms (rows) for each of the feature coordinates while preserving the local and global patterns. For example, the average operator and the maximum operator are the two extremes of the aggregation operation. The average operator sums the atoms together as equals, while the maximum operator broadcasts a single atom. 

The SAF summation operation is given by:
\begin{equation}
f_j = \sum_i^{N_a} P_{ij} \cdot \varepsilon_{ij} \label{eqn:smartagg},
\end{equation}
where $\varepsilon_{ij}$ is the energy of the likelihood of the $i^{th}$ atom to appear in a drug (i.e., positively labeled compound), according to $j^{th}$ feature coordinate, and $P_{ij}$ is a probability-like function that rescales the energy image by comparing atoms. The underlying idea is to amplify local maximum values or minimum values in $\varepsilon$ for each feature coordinate, which otherwise would have been deleted by aggregation. We define the discrete probability function $P_{ij}$ as a Boltzmann distribution:
\begin{equation}
    P_{ij} = \frac{e^{\beta \cdot \varepsilon_{ij}}}{\sum_i^{N_a} e^{\beta \cdot \varepsilon_{ij}}} \label{eqn:mask},
\end{equation}
 For $\beta=1$,  the resulting operator in Eqn.~\ref{eqn:smartagg} is known in the AI (Artificial Intelligence) community as the Softmax operator. $P_{ij}$ has a statistical mechanics interpretation where $\beta$ is one over the temperature. In the 'hot' case, where $\beta = 0$, $P_{ij} = \frac{1}{N_a}$. In this case, the working point is equivalent to performing the average operation. In the 'cold' case, $\beta \rightarrow \infty$, and the resulted operation is $\max(\cdot)$ along the atoms coordinate, and if $\beta \rightarrow -\infty$, then the resulted operation is $\min(\cdot)$ along the atoms coordinate. Hence, the new weighted sum with $P_{ij}$ as weights, extracts only global patterns (among the atoms) when $\beta \rightarrow 0$, and local patterns only when $\beta \rightarrow -\infty,\infty$. For other values of $\beta$, we have intermediate working points, which balance in a unique manner between global and local views. 
 
 In this work, we set $\varepsilon_{ij}$ to be $\alpha_{ij}$. Thus, when $\beta$ is sufficiently large, it follows that the model takes into consideration only a subset of the molecule's atoms. The rest of the atoms are tuned down and have a negligible effect on the final fingerprint. 
 
 For evaluating SAF, the model architecture we used is named D-MPNN (Directional MPNN), with the implementation provided by the Chemprop code distribution (\href{https://github.com/chemprop}{github link}). We replaced the aggregation operator in this pipeline, with the SAF implementation. The generalization error of each model instance was benchmarked, by randomly splitting the dataset into a training set, validation set, and test set ($80\%-10\%-10\%$).

\subsection{iSAF: Interpretability with Smart Aggregation Framework}\label{subsec:isaf}

Interpretability, in terms of finding which atoms are more important for manifesting the targeted property (like antibacterial activity), is crucial for facilitating drug discovery by AI. We define the importance metric, \textit{S}, to be:
\begin{equation}
    S_i = \frac{1}{N_f} \sum_j^{N_f} P_{ij} \qquad i=1,2,..N_a \label{eqn:imetric}
\end{equation}
where $S_i$ is the importance coefficient for the $i^{th}$ atom, and $P_{ij}$ are the probabilities from the SAF operation defined in Eqn.~\ref{eqn:mask}. The logic is that if the value of $S_i$ is steep, then multiple components of the fingerprint vector $\vec{f}$ will be mainly composed of the $i^{th}$ atom encoding (according to Eqn.~\ref{eqn:smartagg}), hence its importance for prediction. It is important to mention that the encodings coming out of the GNN encoder contain only positive entries due to the use of activation functions (specifically ReLU function), so no mutual cancellation can occur between atoms during aggregation. If $\beta=0$, then $S_i = P_{ij} = \frac{1}{N_a}$ for all $i,j$. Meaning, all the atoms correspond to the same importance coefficient. When $\beta>0$,  certain atoms will comply $S_i > \frac{1}{N_a}$ and others $S_i < \frac{1}{N_a}$, creating an order among atoms. This is due to the fact that $\sum_i^{N_a} S_i = \frac{1}{N_f} \sum_i^{N_a} \sum_j^{N_f} P_{ij} = \frac{1}{N_f} \sum_j^{N_f} \sum_i^{N_a} P_{ij} = 1$. The case where $S_i=\frac{1}{N_a}$ for all the atoms (i.e., for all $i$) when $\beta>0$ requires $\varepsilon_{ij} = v_j$ for all $i$ (according to Eqn.~\ref{eqn:mask}, where $v_j\in \mathrm{R}$), which has negligible probability to occur. 

We extend the metric $S_i$ by looking at the transition over $\beta$, i.e., looking at $S_i(\beta)$. A typical discrete matrix $S\in \mathrm{R}^{N_a \times N_{\beta}}$ that contains the importance coefficients, i.e., $S[i,j]=S_i(\beta = F_{\beta} \cdot j)$ (where $F_{\beta}$ is the sampling frequency of $\beta$-axis), should have similar properties to the following example matrix:
\[
S = \begin{pmatrix}
    & \frac{1}{N_a} & > \frac{1}{N_a} & \gg \frac{1}{N_a} & \cdots & \ggg \frac{1}{N_a} \\
    & \frac{1}{N_a} & < \frac{1}{N_a} & \ll \frac{1}{N_a} & \cdots & \sim 0 \\
    & \frac{1}{N_a} & > \frac{1}{N_a} & > \frac{1}{N_a} & \cdots & \gg \frac{1}{N_a} \\
    & \frac{1}{N_a} & \ll \frac{1}{N_a} & \sim 0 & \cdots & \sim 0 \\
    & \vdots & \vdots & \vdots & \cdots & \vdots
\end{pmatrix}
\]
In this example matrix $S$,  part of the atoms (rows) have monotonically increasing importance coefficients with respect to $\beta$ (columns), while other atoms are monotonically decreasing with $\beta$. Since $\sum_i^{N_a} S_i(\beta) = 1$ (as shown before), each column vector in $S$ has the sum of $1$. This means that the monotonic increase in certain atoms' importance is at the expanse of the other atoms. In the results section, we show that the given matrix is representative, specifically the property of monotonicity (Results, section~\ref{subsec:monot}). 

The matrix $S$ is the key for ranking the atoms. Analyzing the right portion of the matrix $S$, yields the most important atoms overall. Analyzing the left portion of $S$, allows to rank the atoms that are less important overall. The distinction between such atoms is lost in steep values of $\beta$. Extracting the trends in the matrix $S$ helps to discard artifacts caused by sub-optimal training. 

To summarize our method, we offer a recipe to interpret molecules in the levels of the atom based on SAF. 
\begin{enumerate}
    \item Train $N_{\beta}$ model-instances for different value of $\beta$ at each time. 
    \item Calculate the importance coefficients $S_i(\beta)$ out of the rescaling mask $P$ from SAF by forward propagation for the targeted molecules.
    \item Rank the atoms according to $S_i(\beta)$ going backwards (from the largest value of $\beta$ to smallest value of $\beta$). First, identify the most important atom, then the second most important atom, and so on. 
\end{enumerate}

\subsection{Training Data}\label{subsec:technicalinfo}

We used a public dataset, containing the effect of $2336$ FDA-approved compounds on the bacteria \textit{E.coli} \cite{yang_analyzing_2019}. From this collection, $120$ compounds exhibited antibiotic activity. 

\section{Results}

\subsection{Using SAF improves antibiotic activity prediction}\label{subsec:auroc}

\begin{figure}[h]
    \centering
    \includegraphics[width=6cm]{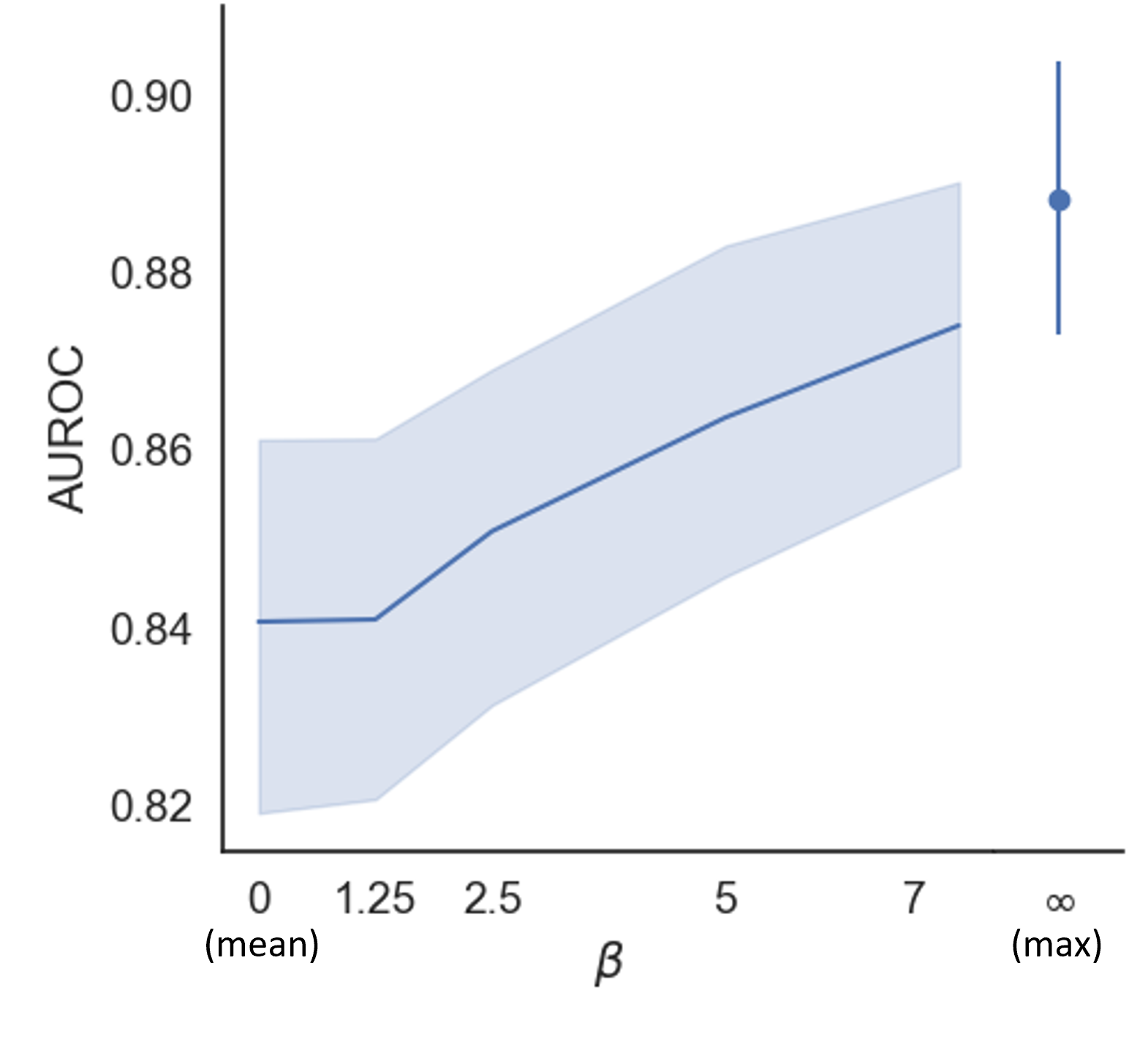}
    \caption{The average AUROC (Area Under the ROC Curve) for predicting antibiotic activity as a function of $\beta$ over $50$ different dataset splits. The error bars are the $95\%$ confidence interval.}
    \label{fig:fig1}
\end{figure}

We evaluated the SAF performances for antibacterial activity prediction. The dataset contains the effect of 2336 FDA-approved compounds on the bacteria \textit{E.coli} (Methods, section ~\ref{subsec:technicalinfo}). We analyzed the effect of SAF with different values of $\beta$ on the ability of D-MPNN to predict antibiotic activity. To achieve a valid evaluation that captures the variability of the process, we perform a valid training over $50$ random unique train-validation-test splits from the dataset for each value of $\beta$. Fig. \ref{fig:fig1} illustrates the AUROC (Area Under the ROC Curve) values received for the converged models over the test sets. 

The results indicate that as $\beta$ increases the AUROC metric increases monotonically. In the antibiotic case, the AUROC increases by $6\%$ between $\beta=0$ and $\beta \rightarrow \infty$ on average. This suggests that increasing the contribution of certain atoms improves the overall performance.

\subsection{Monotonicity of important atoms}\label{subsec:monot}

\begin{figure}[h]
    \centering
    \includegraphics[width=16cm]{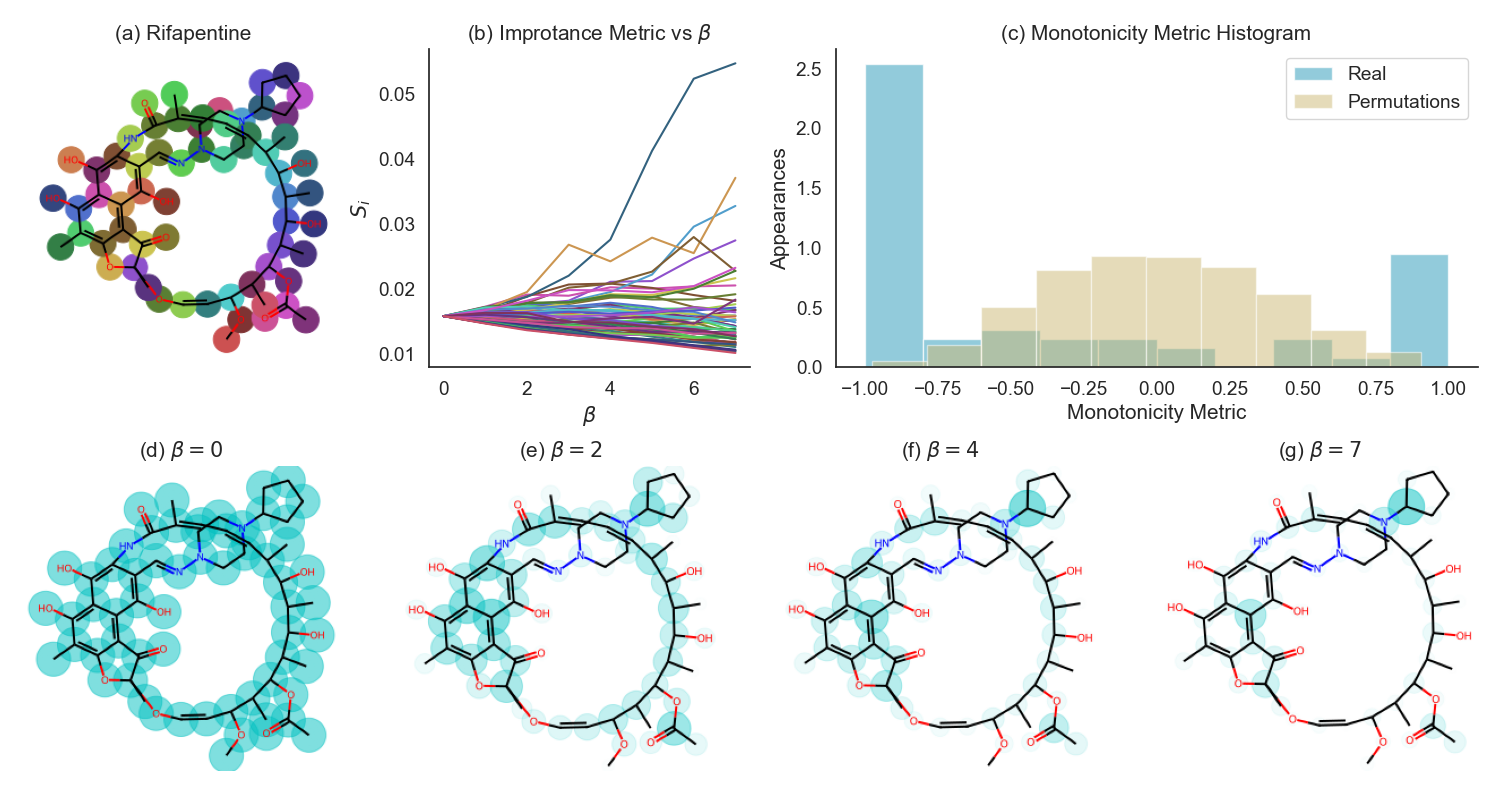}
    \caption{(a) The atomic structure of the antibiotic Rifapentine. (b) The importance coefficients, $S_i$,  as a function of $\beta$ for the different Rifapentine atoms. The colors in (a) and (b) are matching for each atom. (c) Histogram of Spearman's rank correlation coefficient between the $S_i$ in (b) and linear dependence in $\beta$ (cyan bars), and histogram of Spearman's rank correlation coefficient between random permutations of $S_i$ and linear dependence in $\beta$ (yellow). The observed Spearman's correlation distribution is significantly different from random dependence on $\beta$ (p-val$<0.05$, Kolmogorov-Smirnov test).  (d)-(g) Illustration of the smooth and monotonic change in the atom importance. Atomic structure colored by the $S_i$ coefficients (intensity and radius) for different values of $\beta$. }
    \label{fig:fig2}
\end{figure}

The contribution of each atom is captured by $S_i(\beta)$ (Eqn.~\ref{eqn:imetric}). In the case of $\beta=0$, all the atoms have the same weight. As $\beta$ increases, some atoms have higher weights, and for $\beta\rightarrow \infty$ only one atom contributes. The emerging question is whether the important atoms for different values of  $\beta$ are random, or whether the importance order of atoms is kept while increasing $\beta$, thus revealing the importance of atoms gradually. In this section, we show that indeed increasing the hyperparameter $\beta$ reveals in a smooth fashion the atom's importance. Fig. \ref{fig:fig2}-b, illustrates $S_i(\beta)$ for the antibiotic drug  Rifapentine, an example compound. As $\beta$ increases, the atoms' importance quotas emerge in a monotonic and smooth fashion. 

To quantify the monotonicity of the importance metric, we used the Spearman's rank correlation coefficient between $S_i(\beta)$ and a linear function of $\beta$. To show that the results are indeed monotonic, we compared the actual $S_i(\beta)$ coefficients with coefficients that resulted from random permutations between different $\beta$ values. The results indicate a significant monotonicity of $S_i(\beta)$ graphs (Kolmogorov-Smirnov test,  p-val $< 0.05$) (Fig. \ref{fig:fig2}-c). Visually, we can see how the Spearman's rank correlation coefficients in the case of $S_i(\beta)$ are located in the two extreme bins of the histogram, in a significant manner. The meaning of the results is that part of the atoms have monotonically increasing importance graphs, while others regress, which is also visible in Fig.~\ref{fig:fig2}-b, conveying what was stated in Methods, subsection~\ref{subsec:isaf}.

In Fig. \ref{fig:fig2}-d to Fig. \ref{fig:fig2}-g, the atomic structure of the molecule is colored according to the importance pattern given by $\{S_i(\beta)\}_i$ for different values of $\beta$. In the case of $\beta=0$, all the atoms have the same importance coefficient and are colored with the same intensity and radius (Fig.~\ref{fig:fig2}-b,d).  When $\beta$ increases, the importance pattern converges into a subset of  atoms (Fig.~\ref{fig:fig2}-d-g). 

\begin{figure}[h]
    \centering
    \includegraphics[width=17cm]{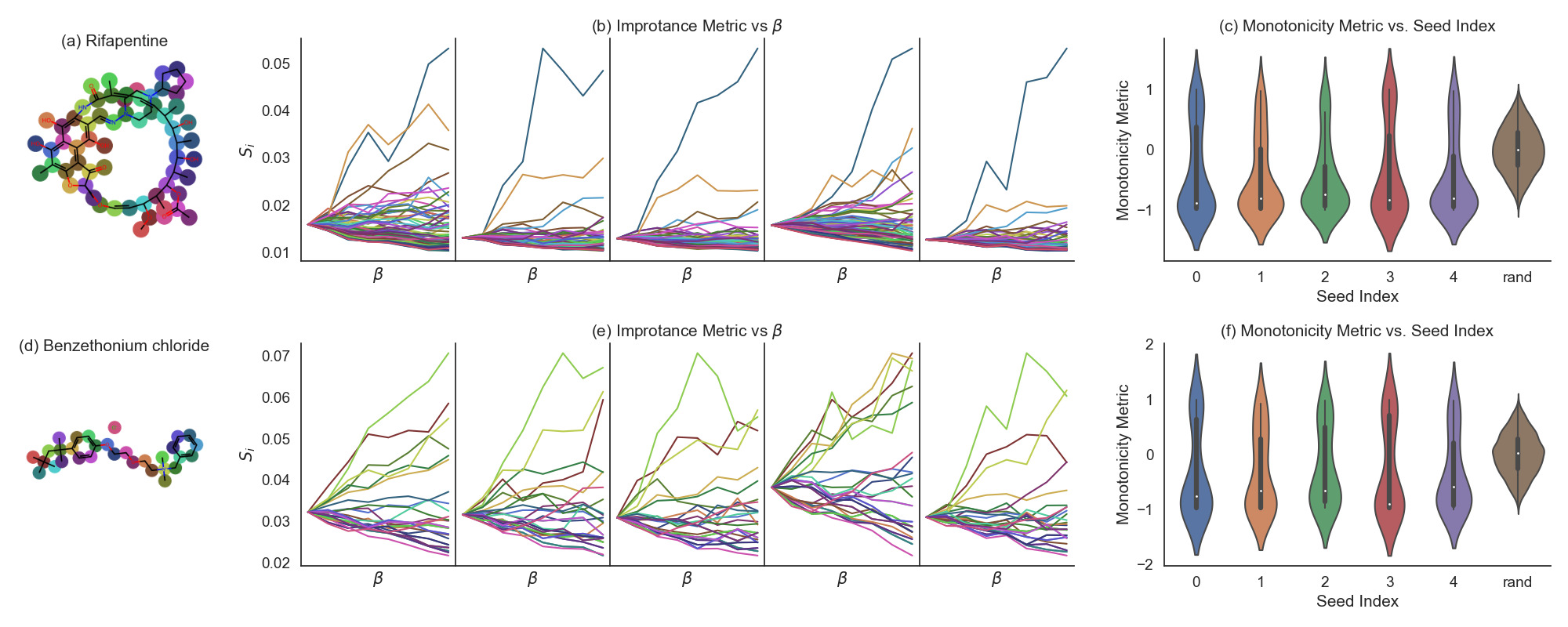}
    \caption{The effect of $\beta$ on the atom importance for five different data splits for  Rifapentine (a) and Benzethonium Chloride (d). (b) The importance coefficients, $S_i$,  as a function of $\beta$. The colors on the atoms' structure are matching to the $S_i$ plots. (c) Violin plot of the distribution of Spearman's rank correlation coefficients between the line plots in (b) and a linear dependence in $\beta$, for $5$ different train-validation-test splits (marked with the seed index). The sixth distribution labeled 'rand' is the distribution of Spearman's rank correlation coefficients between random permutations of the line plots in (b) and a linear dependence in $\beta$. (e)-(f) same as (c)-(d) for Benzethonium Chloride.}
    \label{fig:fig3}
\end{figure}

Fig \ref{fig:fig2}, shows the Rifapentine's results for a single train-validation-test split. Next, we want to show that similar results (with respect to monotonicity) are received for different molecules and dataset splits. To show the monotonicity is not a one-time result of a particular dataset split, we randomly selected $6$ molecules out of the dataset and randomized $5$ train-validation-test splits where all $6$ selected molecules appear in the randomized test sets. So, in total we have $5$ instances for comparison (Fig. \ref{fig:fig3}).  Besides, Rifapentine, Fig. \ref{fig:fig3} also shows the results for  Benzethonium Chloride. The results of all six molecules appear in Figure 1 of the Supplementary Information.

For all the different data splits, there is a significant monotonicity of the atoms' importance (Fig. \ref{fig:fig3}-c and Fig. \ref{fig:fig3}-f).

\subsection{Reproducibility of Atoms Rankings with iSAF}\label{subsec:repo}

\begin{figure}
    \centering
    \includegraphics[width=12cm]{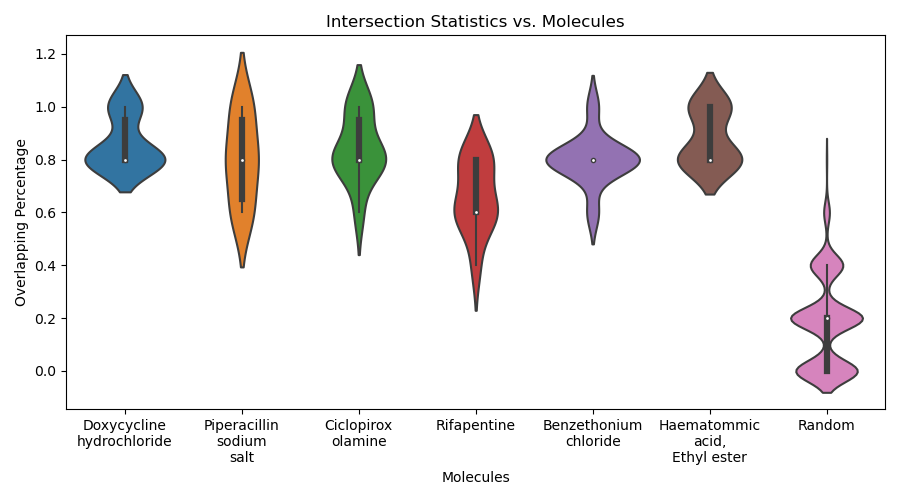}
    \caption{Violin plot of the distributions of the fraction of overlapping atoms among the top-5 most important atoms for $\beta=7$ from different seeds. The last distribution labeled as 'Random' is the distribution of overlapping atoms when the ranking of atoms at each seed is randomized. The overlap distributions for all the molecules are significantly higher than expected by random (p-val$<0.05$, Kolmogorov-Smirnov test).}
    \label{fig:fig4}
\end{figure}

When exploring different dataset splits, a very relevant metric is the size of the overlap between the atoms' orders. That is, the ranking order of atoms does not fluctuate between different $\beta$ values.  To quantify the order's smoothness, we calculated the number of overlapping atoms among the top 5 most important atoms according to $S_i(\beta=7)$ in different seeds. In Fig. \ref{fig:fig4}, we show the distribution of the number of overlapping atoms when comparing between pairs of different train-validation-test splits (for the $6$ molecules in the test set). In Fig. \ref{fig:fig3}-b, \ref{fig:fig3}-e, we can see the overlap visually. 

There is a significant overlap in the top-ranked atoms with an average over $80\%$ ($4$ out of $5$ atoms appear in the intersection). The same can not be said for the random noise. We generated the random samples by permuting the $S_i(\beta)$ sequences (similar to before) and taking the top-5 atoms from the case of $\beta=7$. The resulting p-value is significantly smaller than the $0.05$ mark (Kolmogorov-Smirnov test). Therefore, the importance coefficients $S_i$ are not only very consistent over $\beta$ (due to monotonicity), but also the identity of the important atoms based on them is consistent over different seeds and for different molecules. 

\subsection{iSAF recapitulates $\beta$-Lactam antibiotics functional group}\label{subsec:betalactams}

\begin{figure}[h]
    \centering
    \includegraphics[width=10cm]{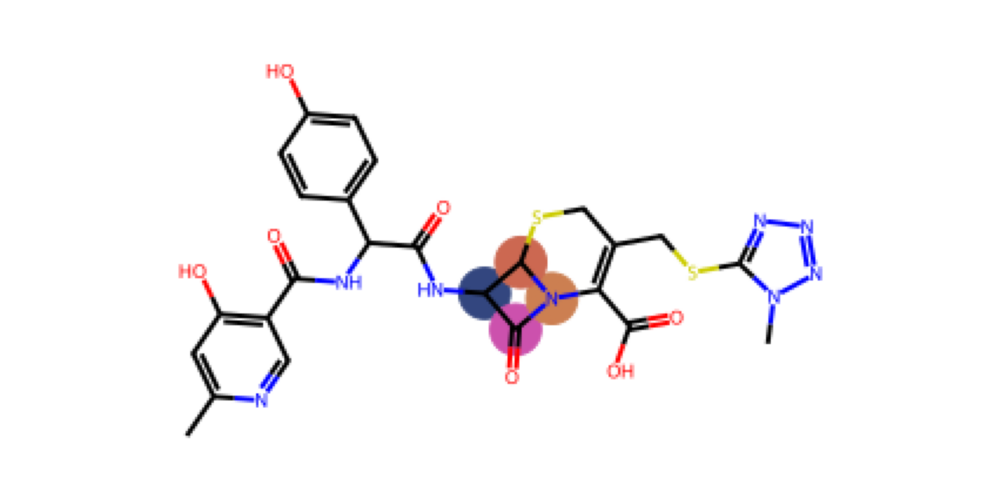}
    \qquad
    \begin{tabular}[b]{|c|c|c|}\hline
        Number of atoms & Probability to be  & Probability to be    \\ examined from  & in  & in $\beta$-Lactam ring \\ 
        the top &  $\beta$-Lactam ring & in random with \\
        & & p-value of 0.05 \\ \hline
        1 & 0.71 & 0.22 \\ \hline 
        2 & 0.91 & 0.35 \\ \hline 
        3 & 0.98 & 0.47 \\ \hline 
        4 & 1 & 0.56 \\ 
        \hline
    \end{tabular}
    \caption{The fraction of $\beta$-lactam antibiotics that at least one of their top-ranked atoms, for $\beta=7$, appears in the $\beta$-Lactam ring. The left column is the number of top atoms considered. For example, for all the 68 $\beta$-Lactam antibiotics, at least one of the 4 top-ranked atoms was within the $\beta$-Lactam ring. The right column contains the fractions of having at least one out of a random set of atoms within the $\beta$-Lactam ring, where the random set is sampled from a random event with probability of $0.05$ (for each molecule).}
    \label{fig:fig5}
\end{figure}

To validate the functional relevance of the emerged importance coefficients, $S_i(\beta)$, we focused on $\beta$-Lactam antibiotics, which include Penicillin derivatives, as an example. These antibiotics have a substructure that accounts for the antibacterial activity, known as the $\beta$-Lactam ring (Fig.~\ref{fig:fig5}). We aimed to validate if the top-ranked atoms, coming out of the $S_i(\beta)$, do reside within the $\beta$-Lactam ring. Hence, we collected all the $\beta$-Lactam antibiotic compounds from test set instances, amounting to $68$ subjects that can be examined. For each compound, we calculated the $S_i(\beta=7)$ out of SAF operation in forward propagation with the model trained on the corresponding train-validation-test split. Then, we counted the number of cases where the top most important atoms reside within the $\beta$-Lactam ring out of the $68$ cases. We did the counting for the first most important atom only, the two most important atoms, the three most important atoms, and the four most important atoms according to $S_i(\beta=7)$. The results are presented in the table in Fig.~\ref{fig:fig5}. 

The results indicate that the representation of the $\beta$-Lactam ring within the important atoms is significantly higher than what is expected randomly. All $68$ cases have at least one significant atom from the top 4 atoms in the $\beta$-Lactam ring. Looking at only the single most important atom in each case, $71\%$ of the cases had it inside the $\beta$-Lactam ring. Interestingly, the atoms from the top that do appear in the ring, are (with high frequency) the same atom in the configuration of the $\beta$-Lactam ring (colored with orange). The bias for the same atom may shed some light on the GNN encoder's view of scaffolds.

\section{Discussion}

One of the main challenges of applying message-passing neural networks for drug discovery is the need to cope with compounds with a different number of atoms. Thus, aggregating the atoms' information is a common bottleneck in many applications.  The common aggregation approach is to sum over all atoms' features, and therefore atoms that are not affected by each other during the message-passing step (which is local) are blended. In this work, we propose a novel aggregating approach where each atom is weighted in a non-linear fashion using the Boltzmann distribution where the 'energies' are the atoms' features from the message-passing stage and the 'temperature' $1/\beta$ is a hyperparameter. When $\beta=0$, the system is 'hot' and the aggregation is a simple sum. When $\beta\rightarrow\infty$, the system is 'cold' and the aggregation is the $\max(\cdot)$ operator. Thus controlling $\beta$ provides a handle over the trade-off between local and global information. 

Our results show that as $\beta$  increases there is an improvement in the antibiotic classification AUROC of 6\% on average. That is, as the network is forced to account for a subset of the overall atoms, the performance improves. Thus, the emerging questions are: What are the groups of atoms that are contributing, and whether is their identity is robust across different cross-validations and different values of $\beta$. 
As $\beta$ increases, fewer atoms contribute significantly to the prediction, yet the \textit{ranking} of the atoms is preserved along different values of $\beta$ (Fig. \ref{fig:fig2}, Fig. \ref{fig:fig3}) and different cross-validation (Fig. \ref{fig:fig4}). That is, the increase of $\beta$ serves as akin of phase separation that amplifies the contribution of the more important atoms in a smooth fashion. Therefore, our approach can be used as a regulated attention mechanism. Unlike other graph attention layers that operate at the message-passing phase by adding more weights that attenuate the message-passing transactions, our approach does not interfere with this phase. This is a crucial point, as it allows for atoms that are  not necessarily in the same neighborhood on the graph, to emerge as significant atoms. 

Another important property of our approach is the ability to reveal the significance at the atom level. To validate our approach, we show that our model highlights the $\beta$-Lactam ring, the functional group of $\beta$-Lactam antibiotics in a statistically significant way. The fact that our schema allows a smooth highlighting of atoms, as $\beta$ increases, provides not only interpretability, but also a better understanding of the dataset composition that could be important for achieving better data-splits. The aggregation operator is considered to be the lesser part of the GNN architecture and is often being neglected. Our work highlights its importance and can be harnessed to other applications of message-passing neural networks.

\begin{acknowledgement}

We thank Tanya Wasserman and the Savir lab members for fruitful discussions. This work is supported by the Israeli Ministry of Science and Technology (MOST) grant \#2149.

\end{acknowledgement}

\bibliographystyle{abbrv}
\bibliography{refs}

\begin{thebibliography}{10}

\bibitem{stephenson_survey_2019}
Natalie Stephenson, Emily Shane, Jessica Chase, Jason Rowland, David Ries, Nicola Justice, Jie Zhang, Leong Chan, and Renzhi Cao.
\newblock Survey of {Machine} {Learning} {Techniques} in {Drug} {Discovery}.
\newblock {\em Current Drug Metabolism}, 20(3):185--193, May 2019.

\bibitem{ekins_exploiting_2019}
Sean Ekins, Ana~C. Puhl, Kimberley~M. Zorn, Thomas~R. Lane, Daniel~P. Russo, Jennifer~J. Klein, Anthony~J. Hickey, and Alex~M. Clark.
\newblock Exploiting machine learning for end-to-end drug discovery and development.
\newblock {\em Nature Materials}, 18(5):435--441, May 2019.

\bibitem{zhang_machine_2017}
Lu~Zhang, Jianjun Tan, Dan Han, and Hao Zhu.
\newblock From machine learning to deep learning: progress in machine intelligence for rational drug discovery.
\newblock {\em Drug Discovery Today}, 22(11):1680--1685, November 2017.

\bibitem{lavecchia_machine-learning_2015}
Antonio Lavecchia.
\newblock Machine-learning approaches in drug discovery: methods and applications.
\newblock {\em Drug Discovery Today}, 20(3):318--331, March 2015.

\bibitem{vamathevan_applications_2019}
Jessica Vamathevan, Dominic Clark, Paul Czodrowski, Ian Dunham, Edgardo Ferran, George Lee, Bin Li, Anant Madabhushi, Parantu Shah, Michaela Spitzer, and Shanrong Zhao.
\newblock Applications of machine learning in drug discovery and development.
\newblock {\em Nature Reviews Drug Discovery}, 18(6):463--477, June 2019.

\bibitem{moriwaki_mordred_2018}
Hirotomo Moriwaki, Yu-Shi Tian, Norihito Kawashita, and Tatsuya Takagi.
\newblock Mordred: a molecular descriptor calculator.
\newblock {\em Journal of Cheminformatics}, 10(1):4, December 2018.

\bibitem{rogers_extended-connectivity_2010}
David Rogers and Mathew Hahn.
\newblock Extended-{Connectivity} {Fingerprints}.
\newblock {\em Journal of Chemical Information and Modeling}, 50(5):742--754, May 2010.

\bibitem{bengio_representation_2014}
Yoshua Bengio, Aaron Courville, and Pascal Vincent.
\newblock Representation {Learning}: {A} {Review} and {New} {Perspectives}, April 2014.
\newblock arXiv:1206.5538 [cs].

\bibitem{duvenaud_convolutional_2015}
David Duvenaud, Dougal Maclaurin, Jorge Aguilera-Iparraguirre, Rafael Gómez-Bombarelli, Timothy Hirzel, Alán Aspuru-Guzik, and Ryan~P. Adams.
\newblock Convolutional {Networks} on {Graphs} for {Learning} {Molecular} {Fingerprints}, November 2015.
\newblock arXiv:1509.09292 [cs, stat].

\bibitem{lin_kgnn_2020}
Xuan Lin, Zhe Quan, Zhi-Jie Wang, Tengfei Ma, and Xiangxiang Zeng.
\newblock {KGNN}: {Knowledge} {Graph} {Neural} {Network} for {Drug}-{Drug} {Interaction} {Prediction}.
\newblock In {\em Proceedings of the {Twenty}-{Ninth} {International} {Joint} {Conference} on {Artificial} {Intelligence}}, pages 2739--2745, Yokohama, Japan, July 2020. International Joint Conferences on Artificial Intelligence Organization.

\bibitem{li_grapher_2020}
Bing Li, Wei Wang, Yifang Sun, Linhan Zhang, Muhammad~Asif Ali, and Yi~Wang.
\newblock {GraphER}: {Token}-{Centric} {Entity} {Resolution} with {Graph} {Convolutional} {Neural} {Networks}.
\newblock {\em Proceedings of the AAAI Conference on Artificial Intelligence}, 34(05):8172--8179, April 2020.

\bibitem{yang_analyzing_2019-1}
Kevin Yang, Kyle Swanson, Wengong Jin, Connor Coley, Philipp Eiden, Hua Gao, Angel Guzman-Perez, Timothy Hopper, Brian Kelley, Miriam Mathea, Andrew Palmer, Volker Settels, Tommi Jaakkola, Klavs Jensen, and Regina Barzilay.
\newblock Analyzing {Learned} {Molecular} {Representations} for {Property} {Prediction}.
\newblock {\em Journal of Chemical Information and Modeling}, 59(8):3370--3388, August 2019.

\bibitem{stokes_deep_2020-1}
Jonathan~M. Stokes, Kevin Yang, Kyle Swanson, Wengong Jin, Andres Cubillos-Ruiz, Nina~M. Donghia, Craig~R. MacNair, Shawn French, Lindsey~A. Carfrae, Zohar Bloom-Ackermann, Victoria~M. Tran, Anush Chiappino-Pepe, Ahmed~H. Badran, Ian~W. Andrews, Emma~J. Chory, George~M. Church, Eric~D. Brown, Tommi~S. Jaakkola, Regina Barzilay, and James~J. Collins.
\newblock A {Deep} {Learning} {Approach} to {Antibiotic} {Discovery}.
\newblock {\em Cell}, 180(4):688--702.e13, February 2020.

\bibitem{velickovic_graph_2018}
Petar Veličković, Guillem Cucurull, Arantxa Casanova, Adriana Romero, Pietro Liò, and Yoshua Bengio.
\newblock Graph {Attention} {Networks}, February 2018.
\newblock arXiv:1710.10903 [cs, stat].

\bibitem{islam_mpool_2023}
Muhammad Ifte~Khairul Islam, Max Khanov, and Esra Akbas.
\newblock {MPool}: {Motif}-{Based} {Graph} {Pooling}, March 2023.
\newblock arXiv:2303.03654 [cs].

\bibitem{ju_tgnn_2023}
Wei Ju, Xiao Luo, Meng Qu, Yifan Wang, Chong Chen, Minghua Deng, Xian-Sheng Hua, and Ming Zhang.
\newblock {TGNN}: {A} {Joint} {Semi}-supervised {Framework} for {Graph}-level {Classification}, April 2023.
\newblock arXiv:2304.11688 [cs].

\bibitem{ying2019gnnexplainer}
Rex Ying, Dylan Bourgeois, Jiaxuan You, Marinka Zitnik, and Jure Leskovec.
\newblock Gnnexplainer: Generating explanations for graph neural networks, 2019.

\bibitem{vinas_hypergraph_2023}
Ramon Viñas, Chaitanya~K. Joshi, Dobrik Georgiev, Phillip Lin, Bianca Dumitrascu, Eric~R. Gamazon, and Pietro Liò.
\newblock Hypergraph factorization for multi-tissue gene expression imputation.
\newblock {\em Nature Machine Intelligence}, 5(7):739--753, July 2023.

\bibitem{yang_analyzing_2019}
Kevin Yang, Kyle Swanson, Wengong Jin, Connor Coley, Philipp Eiden, Hua Gao, Angel Guzman-Perez, Timothy Hopper, Brian Kelley, Miriam Mathea, Andrew Palmer, Volker Settels, Tommi Jaakkola, Klavs Jensen, and Regina Barzilay.
\newblock Analyzing {Learned} {Molecular} {Representations} for {Property} {Prediction}.
\newblock {\em Journal of Chemical Information and Modeling}, 59(8):3370--3388, August 2019.

\end{thebibliography}

\end{document}